\documentclass[prl,aps,preprint,nobalancelastpage]{revtex4}

\usepackage{graphicx}
\usepackage{dcolumn}
\usepackage{bm}

\begin{document}

\title{Non-generality of the Kadowaki-Woods ratio in correlated oxides}

\author{N. E. Hussey}
\affiliation{ H. H. Wills Physics Laboratory, University of Bristol, Tyndall Avenue, BS8 1TL, United Kingdom}%

\date{today}

\begin{abstract}

An explicit expression for the Kadowaki-Woods ratio in correlated metals is derived by invoking saturation of the (high-frequency)
Fermi-liquid scattering rate at the Mott-Ioffe-Regel limit. Significant deviations observed in a number of oxides are quantitatively
explained due to variations in carrier density, dimensionality, unit cell volume and the number of individual sheets in the
Brillouin zone. A generic re-scaling of the original Kadowaki-Woods plot is also presented.

\end{abstract}

\maketitle

One of the triumphs of Landau-Fermi-liquid theory is its ability to encapsulate the physical behavior of a
wide variety of metals with only a limited set of parameters that characterize the many-body enhancement of the (quasiparticle) effective mass.
This enhancement manifests itself in a number of physical properties including the magnetic susceptibility $\chi_P$, the electronic
specific heat coefficient $\gamma_0$, the coefficient $A$ of the $T^2$ resistivity and the slope of the low-$T$ thermoelectric
power $S$/$T$. Important empirical relations correlating these quantities have been found, including the Kadowaki-Woods ratio (KWR)
($A$/$\gamma_0^2$ $\sim$ $a_0$ = 10$^{-5}$ $\mu\Omega$cm.mol$^2$.K$^2$/J$^2$) \cite{kwr}, the Wilson ratio
($\chi_P$/$\gamma_0$ $\sim$ 1 T$^2$/K$^2$ for strongly-correlated fermions) \cite{wilson} and most recently, the
Behnia-Jaccard-Flouquet ratio ($S$/$\gamma_0$$T$ $\sim$ 10$^5$ C/mol) \cite{behnia}.

In the original KWR paper \cite{kwr}, only heavy fermion metals containing U and Ce were considered and most theoretical treatments
to date have focussed on heavy fermions and the dependence of $A$ on the $f$-electron density of states \cite{yamada,levin,coleman,miyake,kontani}.
Since then however, the ratio has been examined in a broad range of correlated metals and whilst the perception is one of generality,
there are some notable exceptions, particularly among the oxides. As illustrated in Fig. 1a for example, La$_{1.7}$Sr$_{0.3}$CuO$_4$ \cite{nakamae},
Ca$_{2-x}$Sr$_x$RuO$_4$ \cite{maeno,nakatsuji}, La$_{1-x}$Sr$_x$TiO$_3$ ($x$ $\agt$ 0.9) \cite{okuda}, LiV$_2$O$_4$ \cite{urano}, V$_2$O$_3$ and
Na$_{0.7}$CoO$_2$ \cite{li} all show significant deviations from the empirical line $A$/$\gamma_0^2$ $\sim$ $a_0$. In this Letter, we show that
whilst the KWR is largely insensitive to the strength of electron correlations (as manifest in the mass renormalization), its value is strongly
material-specific. Deviations from $a_0$ {\it for all those oxides listed above} are qualitatively and quantitatively explained with a minimum
of assumptions. Though we focus here on oxides, our approach is sufficiently general as to encompass all correlated metals, including heavy fermions,
and a generic revision of the original KW scaling is proposed.

For a highly correlated Fermi-liquid, one can neglect electron-phonon scattering and define a $T$- and $\omega$-dependent scattering rate
of the form $\Gamma_{\rm{FL}}$($T$, $\omega$) = $\Gamma_0$ + $\Phi$(($p \pi T$)$^2$ + $\omega^2$), where $\Gamma_0$ is the impurity scattering
rate, $\Phi$ is a coefficient to be determined and $p$ = 2 for electron-electron scattering \cite{gurzhi}. This form for $\Gamma_{\rm{FL}}$
reflects the phase space available for scattering and is appropriate at low energies. At higher energies however, the scattering
rate must approach some maximum or `saturated' value of order the bare bandwidth $W$. To account for this, we introduce a maximum scattering rate
$\Gamma_{\rm{max}}$ = $v_F$/$a$ (where $v_F$ is the unrenormalized Fermi velocity and $a$ the lattice spacing) that is compatible with the
Mott-Ioffe-Regel limit, and define an {\it effective} transport scattering rate $\Gamma_{\rm{eff}}$($T$, $\omega$) via
\begin{equation}
\frac{1}{\Gamma_{\rm{eff}}(T, \omega)} = \frac{1}{\Gamma_{\rm{FL}}(T, \omega)} + \frac{1}{\Gamma_{\rm{max}}}\label{1}
\end{equation}

Inserting (\ref{1}) into the Drude formula for the dc conductivity, one obtains the well-known parallel-resistor formula for saturating
metals \cite{hussey1}. Moreover, by including basal plane anisotropy in $\Gamma_{\rm{eff}}$($T$, $\omega$), this form of scattering
rate can successfully account for both the dc and optical transport properties of optimally doped cuprates \cite{hussey2,hussey3}.
To simplify our working, we make the identity $\Lambda$ = $\Gamma_{\rm{max}}$ + $\Gamma_{\rm{0}}$ + (2$\pi T$)$^2$ and re-arrange
$\Gamma_{\rm{eff}}$ to give $\Gamma_{\rm{eff}}$ = $\Gamma_{\rm{max}}$ - ($\Gamma_{\rm{max}}$/$\Lambda$)(1/(1 + $\Phi
\omega^2$/$\Lambda$)). In this form, $\lambda_{tr}$($T$, $\omega$) can be obtained analytically from the appropriate Kramers-Kronig (KK)
transformation \cite{puchkov} with
\begin{equation}
\lambda_{tr}(T, \omega) = \Gamma_{\rm{max}}^2 (\frac{\Phi}{\Lambda})^{1/2}  \frac{1}{(\Lambda + \Phi \omega^2)}\label{2}
\end{equation}
Extrapolating to $\omega$ = 0 and low $T$ where $\Lambda$ $\sim$ $\Gamma_{\rm{max}}$, we finally arrive at our expression for the
dc mass enhancement factor, $\lambda_{tr}$(0) = ($\Phi \Gamma_{\rm{max}}$)$^{1/2}$ = ($\Phi v_F$/$a$)$^{1/2}$. Note that this mass
enhancement is relative to the band mass $m_b$ as would be calculated, say, from LDA band calculations, and not the bare electron mass
$m_0$. Finally, by converting to resistivity $\rho$($T$) = $\rho_0$ + $AT^2$ = $m_b$/$ne^2$($\Gamma_0$ + (2$\pi T$)$^2$) and re-instating
all parameters, we obtain
\begin{equation}
A = \frac{4 \pi^2 k_B^2}{e^2 \hbar^2} \frac{a m_b}{n}  (\frac{\lambda_{tr}(0)^2}{v_F})\label{3}
\end{equation}
Note that $A$ is proportional to $\lambda_{tr}$(0)$^2$ but also depends on $\Gamma_{\rm{max}}$ ($v_F$/$a$). This somewhat surprising result
can be understood by acknowledging that $\Gamma_{\rm{max}}$ sets the scale for $\Gamma_{\rm{eff}}$($\omega$) to vary between its
low- and high-frequency limits. This in turn, via the KK transformation, determines the enhancement in $\lambda_{tr}$(0) as $\omega$ $\rightarrow$ 0.
$\Gamma_{\rm{max}}$ has in fact appeared in several previous derivations of $A$ \cite{coleman,miyake,lirasul} though not in this form.
Miyake {\it et al.} for example derived an expression for the KWR in heavy fermions assuming a strong frequency dependence of the
{\it quasi-particle} lifetime $\Sigma$($\omega$) with saturation at the unitary limit \cite{miyake}. Their treatment of saturation however
did not allow for an analytical derivation of $\lambda_{tr}$(0) and the resulting expression for $A$ was markedly different from that given
in (\ref{3}).

The electronic specific heat coefficient $\gamma_V$ = 1/3 $\pi^2 k_B^2$ (1 + $\lambda_{th}$) $\int$ d$S$/4$\pi^3 \hbar v_F$
where (1 + $\gamma_{th}$) contains all contributions to the thermodynamic mass enhancement $m^*$/$m_b$, {\it including} electron-phonon
coupling $\lambda_{ph}$. Thus provided $\lambda_{tr}$(0) $\sim$ (1 + $\lambda_{th}$) $\gg$ $\lambda_{ph}$, correlation effects will
cancel in the ratio $A$/$\gamma_V^2$ and the empirical scaling of the KWR in correlated metals is obtained. (Conversely when $\lambda_{tr}$(0)
$\ll$ 1, the KWR will be substantially reduced.) For intermediate $\lambda_{tr}$(0) ($m^*$ $\sim m_b$), the KWR will depend
sensitively on knowledge of $m_b$ and thus on the accuracy of the band calculations. In the following therefore, we choose to ignore
the ratio ($\lambda_{tr}$(0)/(1 + $\lambda_{th}$))$^2$, acknowledging that in most cases, this will lead to an overestimate of the KWR.
With this in mind, we now consider factors that might influence the KWR and try to quantify their impact on those oxides listed above.

{\it Effect of unit cell volume}: Ever since Kadowaki and Woods' seminal paper \cite{kwr}, it has become customary to plot the KWR with $A$
expressed in units of $\mu\Omega$cm/K$^2$ and $\gamma_0$ in J/mol.K$^2$ (or mJ/mol.K$^2$), as illustrated in Fig. 1a. The electrical conductivity
$\sigma$ of a metal is the response function describing a current density {\bf J} that is in turn related to a carrier density $n$
expressed in units of m$^{-3}$. Hence, in its original form, the KWR compares a {\it volume} quantity ($A$) with a {\it molar} quantity
($\gamma_0$). In order to compare the two quantities directly, we suggest it is more appropriate to express $\gamma_0$ in its volume form
$\gamma_V$ as given above. The two are scaled by the ratio $N_A V$/$Z$ where $N_A$ is Avogadro's number, $V$ is the unit cell volume and
$Z$ the number of formula units per unit cell.

This seemingly mute point, the choice of units, can have dramatic consequences. In the layered cobaltate Na$_{0.7}$CoO$_2$ for example,
$A$/$\gamma_0^2$ $\sim$ 50 $a_0$, {\it almost two orders of magnitude larger} than that seen in heavy-fermions \cite{li} (see Fig.\ref{1}a).
This remarkable enhancement was naturally viewed as a signature of intense electron-electron scattering, possibly arising from
magnetic frustration in the triangular lattice or proximity to a quantum critical point. Significantly however, Na$_{0.7}$CoO$_2$ has a tiny
unit cell (hcp lattice, $a$ = 2.84$\AA$, $c$ = 10.94$\AA$, $V$ = 76$\AA^3$ and $Z$ = 2). By contrast, in La$_{1.7}$Sr$_{0.3}$CuO$_4$, where
$Z$ = 1 in a unit cell is of comparable size (bct lattice, $a$ = 3.86$\AA$, $c$ = 6.4$\AA$, $V$ = 95$\AA^3$), $A$/$\gamma_0^2$ $\sim$ 5$a_0$
\cite{nakamae}. If we now define a new parameter for the KWR, $b_0$ = 1 $\mu\Omega$cmK$^2$.cm$^6$/J$^2$, we find for Na$_{0.7}$CoO$_2$,
$A$/$\gamma_V^2$ = 0.29$b_0$ while for La$_{1.7}$Sr$_{0.3}$CuO$_4$, $A$/$\gamma_V^2$ = 0.17$b_0$. Hence, the one order of magnitude difference
in the two original KWR values can be attributed largely to the factor ($V$/$Z$)$^2$. Fig.\ref{1}b shows our revision
of the KW plot in which $A$ is compared with $\gamma_V$ rather than $\gamma_0$. Note that the KWR for both V$_2$O$_3$ \cite{li} and
LiV$_2$O$_4$ \cite{urano} are also strongly renormalized in this new scaling plot. These striking results serve to underline the importance of
units, particularly when comparing compounds of very different chemical composition.

{\it Effect of dimensionality}: The dashed line in Fig. 1b corresponds to a nominal KWR, $A$/$\gamma_V^2$ = 0.2$b_0$. All compounds near this
line are quasi-two-dimensional (quasi-2D) metals whose physics is dominated by a single (large) cylindrical Fermi surface (FS) (the one exception
being La$_{0.05}$Sr$_{1.95}$TiO$_3$ to be discussed in the following section). Those compounds found below this line have either closed or multiple FS
or a combination of the two. We note that whilst $A$ depends on the FS {\it volume} (through $n$), $\gamma_V$ is largely governed by the FS {\it area}.
Thus we expect the KWR to be sensitive to the FS geometry. Table 1 summarizes our derived KWR for spherical (3D), cylindrical (2D) and planar
(1D) FS. Note that once we ignore the correlation term, there are {\it no} adjustable parameters in these expressions.

\begin{table}
\caption{$A$/$\gamma_V^2$ for spherical (3D), cylindrical (2D) and planar (1D) Fermi surfaces.}
\begin{ruledtabular}
\begin{tabular}{cc}
Fermi surface&$A$/$\gamma_V^2$ (10$^{14} \mu \Omega$cm/K$^2$)/(mJ/cm$^3$/K$^2$)$^2$\\
\hline 3D&(108 $\pi^4 \hbar$/$e^2 k_B^2$) ($a$/$k_F^6$) ($\lambda_{tr}^2$(0)/(1 + $\lambda_{th}$)$^2$)\\
\hline 2D&(72 $\pi \hbar$/$e^2 k_B^2$) ($a c^3$/$k_F^3$) ($\lambda_{tr}^2$(0)/(1 + $\lambda_{th}$)$^2$)\\
\hline 1D&(9 $\pi \hbar$/2$e^2 k_B^2$) $a b^3 c^3$ ($\lambda_{tr}^2$(0)/(1 + $\lambda_{th}$)$^2$)\\
\end{tabular}
\end{ruledtabular}
\label{table}
\end{table}

In order to compare directly with the KWR of real materials, detailed FS information is required. The FS of both Na$_{0.7}$CoO$_2$ and
La$_{1.7}$Sr$_{0.3}$CuO$_4$ is found to be approximately cylindrical with radii of $k_F$ = 0.65 and 0.55$\AA^{-1}$
respectively \cite{hasan,ino}. Inserting these values into our 2D expression for the KWR, we find $A$/$\gamma_V^2$ = 0.66$b_0$ for Na$_{0.7}$CoO$_2$
and 0.3$b_0$ for La$_{1.7}$Sr$_{0.3}$CuO$_4$. Thus, the enhanced KWR in both compounds can be adequately explained by consideration of the
combined effects of dimensionality and unit cell volume, {\it without the need to invoke additional or exotic scattering}.

{\it Effect of carrier density}: From Table 1 we see that the KWR in 1D metals is independent of $k_F$, though not the unit cell dimensions.
(Because of this, one expects the KWR to be extremely large in 1D organics). In 2D and 3D systems however, the KWR depends strongly
on $k_F$. An ideal material to test this relation is La$_{1-x}$Sr$_x$TiO$_3$ for which $n$ changes continuously for 0 $\alt$ $x$ $\alt$ 1. At $x$ = 1,
the system is close to being a band insulator, whilst for $x$ $\alt$ 0.05, it is a Mott insulator. In between, the system exhibits metallic transport
characterized by a large $T^2$ resistivity that diverges at both ends of the series \cite{okuda,tokura}.

The inset in Fig. 2 shows the KWR for La$_{1-x}$Sr$_x$TiO$_3$ near $x$ = 0 (closed circles, reproduced from Ref. \cite{tokura}). As indicated by
the dashed line, $A$ is NOT proportional to $\gamma_V^2$. According to Table 1, one must also take into account the variation in $n$. If we assume
the FS in La$_{1-x}$Sr$_x$TiO$_3$ to be spherical, we can write 1/$k_F^6$ = 1/(3$\pi^2 n$)$^2$ and thus we expect $A$/$\gamma_V^2$
to be proportional to 1/$n^2$ (the constant of proportionality here being $m$(3$\hbar$/$e^2 k_B^2$)$a$ where $m$ accounts for the presence of
multiple bands - see below). By re-plotting the data as $An^2$ versus $\gamma_V^2$ (black squares in the inset), linear scaling is indeed
recovered. In the main panel, $A$/$\gamma_V^2$ is plotted versus $a$/$n^2$ for a range of $x$ values between 0 and 1 \cite{okuda,tokura}.
The dashed line is the best fit through the data set, the slope being $m$ = 0.5. Remarkably, the scaling appears to hold across
the entire series with $A$/$\gamma_V^2$ {\it varying by 5 orders of magnitude}. Previous derivations of the KWR have contained some dependence
on carrier number \cite{yamada,miyake,kontani} but never as strong as that shown in Fig. 2. The persistence of KW scaling
towards $x$ = 1 is somewhat surprising, but does suggest that electron correlations continue to play a prominent role in the
low-$T$ transport behavior in La$_{1-x}$Sr$_x$TiO$_3$ right across the series.

{\it Multiple band effects}: Significant deviations from the KWR are also expected when several bands cross the Fermi level or when a single band
is split into individual sheets. The key point here is that whilst bands contribute `in series' to $\gamma_V$, they add `in parallel' to $A$.
Obviously, when bands have different sizes and masses, the problem is rather complicated. Provided these are known however, one can in principle obtain a
quantitative estimate for $A$/$\gamma_V^2$. To illustrate this point, we consider Sr$_2$RuO$_4$, perhaps the best characterized multi-band oxide.
The FS of Sr$_2$RuO$_4$ comprises three cylinders ($\alpha$, $\beta$ and $\gamma$) formed from 4$t_{2g}$ orbitals in the RuO$_2$ planes.
The $k_F$ and $m^*$ values are 0.3, 0.62 and 0.75($\AA^{-1}$) and 3.3, 7.0 and 16.0 $m_0$ for $\alpha$, $\beta$ and $\gamma$ respectively
\cite{bergemann} while $A$ = 4.5 - 7.5 n$\Omega$cm/K$^2$, $\gamma_V$ = 0.66 mJ/cm$^3$.K$^2$ and $A$/$\gamma_V^2$ = 0.01 - 0.015$b_0$ \cite{maeno},
i.e. more than one order of magnitude smaller than in Na$_{0.7}$CoO$_2$ and La$_{1.7}$Sr$_{0.3}$CuO$_4$. Note that a similar KWR is found in CaVO$_3$
($A$/$\gamma_V^2$ = 0.011$b_0$ \cite{inoue1}), whose FS has three inter-penetrating cylinders \cite{inoue2}.

The specific heat is most easily dealt with by re-writing the expression for $\gamma_V$ in terms of $m^*$, i.e. $\gamma_V$ =
($\pi k_B^2$/3$\hbar^2 c$)$\Sigma_i$$m_i^*$. Inserting the above masses, one finds $\gamma_V$ = 0.67 mJ/cm$^3$.K$^2$, in excellent agreement with experiment.
From (\ref{3}) (and assuming $m^*$ $\gg$ $m_b$), the $A$ coefficient for a single 2D cylinder is $A_i \sim$ (8$\pi^3 a c k_B^2$/$e^2 \hbar^3$).($m_i^{*2}$/$k_F^3$),
from which we obtain $A_{\alpha}$ = 12.4, $A_{\beta}$ = 6.4 and $A_{\gamma}$ = 15.2 n$\Omega$cm/K$^2$. In order to estimate the magnitude of the
combined $A$ coefficient, we must assume that each sheet acts as an independent conduction channel. When the $A_i$ coefficients are very large
compared to $\rho_0$ (in the relevant temperature range), one can simply apply the parallel-resistor formula, i.e. 1/$A$ = $\Sigma_i$1/$A_i$
= 3.6 n$\Omega$cm/K$^2$. In the opposite limit ($\rho_0$ $\gg$ $AT^2$), the weighting of individual contributions to
$\rho_0$ should also be taken into account via (see Appendix)
\begin{equation}
A = \frac{A_{\alpha} \rho_{0\beta}^2 \rho_{0\gamma}^2 + A_{\beta} \rho_{0\alpha}^2 \rho_{0\gamma}^2 + A_{\gamma} \rho_{0\alpha}^2 \rho_{0\beta}^2}
{(\rho_{0\alpha} \rho_{0\beta} + \rho_{0\alpha} \rho_{0\gamma} + \rho_{0\beta} \rho_{0\gamma})^2}\label{4}
\end{equation}
where $\rho_{0i}$ = ($\Sigma_i$$k_{Fi}$)/$k_{Fi}$ for a 2D metal. Eqn. (\ref{4}) gives $A$ = 5.2 n$\Omega$cm/K$^2$ for Sr$_2$RuO$_4$. Both
estimates are comparable and agree well with experiment.

As an independent test of this picture, we consider Ca$_{2-x}$Sr$_x$RuO$_4$. For $x$ $\alt$ 0.5, quasi-particles on the $\alpha$ and $\beta$ bands
tend to localize, leaving only itinerant (and extremely heavy) quasiparticles on the large $\gamma$ band \cite{nakatsuji}. At $x$ = 0.2,
$A$/$\gamma_V^2$ = 0.18$b_0$ \cite{nakatsuji}. Applying our single-band (2D) expression
from Table 1 to Ca$_{1.8}$Sr$_{0.2}$RuO$_4$ (and assuming no change in the size of the $\gamma$-sheet), we obtain $A$/$\gamma_V^2$ = 0.12$b_0$.
Hence, the very different KWR in the two ruthenates can be qualitatively and quantitatively understood by acknowledging the transition
from multi- to single band physics with Ca doping. Proximity to the Mott insulating state is seen to induce negligible enhancement in $A$/$\gamma_V^2$.

In summary, we have derived explicit expressions for the KWR in correlated metals in which mass renormalization is effectively redundant.
Deviations from the original KWR in a host of correlated oxides have been explained by careful consideration of the unit cell volume, dimensionality,
carrier density and multi-band effects. Moreover, the importance of using appropriate units in plotting the KWR has been aptly demonstrated.
Though independent estimates of $p$ (e.g. from optical conductivity) and a full microscopic derivation of Eqn. (\ref{3}) are required, the overall
consistency with experiment suggests that our assumption in (\ref{1}) is valid and our expression for $A$ may be used to gain additional information
on the underlying physics in a variety of compounds.

When extending this scheme to other systems, additional effects, such as disorder or orbital degeneracy (thought to play a key role in
Yb-based compounds for example \cite{kontani,tsujii}), should also be taken into account. In the light of all these complications,
it is perhaps worth commenting on the perceived generality of the KWR, especially in heavy fermions. Though heavy fermions are mostly 3D compounds,
$\lambda_{tr}$(0) $\gg$ 1 and the unit cell is uniformly large, the Fermi surfaces are complicated with numerous sheets of varying size
and structure. Thus, their adherence to the KW scaling appears somewhat puzzling. In order to reconcile this within the suggested framework,
one must assume both $A$ and $\gamma_V$ are dominated by a single surface (of heavy mass). Only when full FS information is available
(i.e. that can account for the {\it entire} $\gamma_V$) however, can the KWR be calculated for individual materials. We therefore reserve a
full discussion on the KWR in heavy fermions for a later date.

Finally, in line with Luttinger's theorem, we expect the KWR to remain constant as one varies $W$ (but not $n$) and approach the Mott insulating state
from the metallic side. This is supported by our quantitative explanation of the KWR in Ca$_{1.8}$Sr$_{0.2}$RuO$_4$. We argue that only when our revised
form of the KWR is used (i.e. with the appropriate units), can genuine departures from the empirical scaling law, e.g near a quantum critical point,
be taken as evidence of novel physics. We hope this work stimulates a more rigorous approach to the physics of correlated metals and we welcome
further quantitative comparisons on other systems in due course.

The author would like to thank J. C. Alexander, K. Behnia, A. Fujimori, K. Kadowaki, H. Kontani, Y. Matsuda and H. Takagi for stimulating and
enlightening discussions and EPSRC for their support. The author also acknowledges the University of Tokyo for their hospitality during the course of this work.

\appendix*
\subsubsection{$A$ coefficient in a 3-band quasi-2D metal}
The total conductivity  $\sigma_T$ = $\sigma_{\alpha}$ + $\sigma_{\beta}$ + $\sigma_{\gamma}$ is taken as the sum of individual contributions from
each band. At 0K, one can assume an isotropic-$\ell$ approximation and write,
\begin{equation}
\sigma_{0T}= \frac{e^2}{2\pi \hbar c} \ell_0 \Sigma_i k_{Fi} = \frac{1}{\rho_{0T}} =
\frac{\rho_{0\alpha} \rho_{0\beta} + \rho_{0\alpha} \rho_{0\gamma} + \rho_{0\beta} \rho_{0\gamma}}{\rho_{0\alpha} \rho_{0\beta} \rho_{0\gamma}}\label{A.1}
\end{equation}
At finite temperature, $\sigma_i$ = 1/($\rho_{0i}$ + $A_i$$T^2$) and so the change in conductivity is given by
$\Delta \sigma_i$ = $\sigma_i$ - $\sigma_{0i}$ = -$A_i T^2$/$\rho_{0i}$($\rho_{0i}$ + $A_i T^2$) $\sim$ -$A_i T^2$/$\rho_{0i}^2$
provided $\rho_{0i}$ $\gg$ $A_i$$T^2$. Thus,
\begin{equation}
\Delta \sigma_T = -(\frac{A_{\alpha}}{\rho_{0\alpha}^2} + \frac{A_{\beta}}{\rho_{0\beta}^2} + \frac{A_{\gamma}}{\rho_{0\gamma}^2})T^2
= -(\frac{A_{\alpha} \rho_{0\beta}^2 \rho_{0\gamma}^2 + A_{\beta} \rho_{0\alpha}^2 \rho_{0\gamma}^2 + A_{\gamma} \rho_{0\alpha}^2 \rho_{0\beta}^2}
{(\rho_{0\alpha} \rho_{0\beta} \rho_{0\gamma})^2})T^2\label{A.3}
\end{equation}
Since $\Delta\sigma_T$/$\sigma_{0T}$ = - $\Delta\rho_T$/$\rho_{0T}$, the total change in {\it resistivity} $\Delta\rho_T$ = $AT^2$ with
$A$ as given in (\ref{4}). This can of course be generalized to an $n$-band metal or to other dimensions.

\newpage
\noindent
\begin{figure}
\includegraphics[width=8.0cm,keepaspectratio=true]{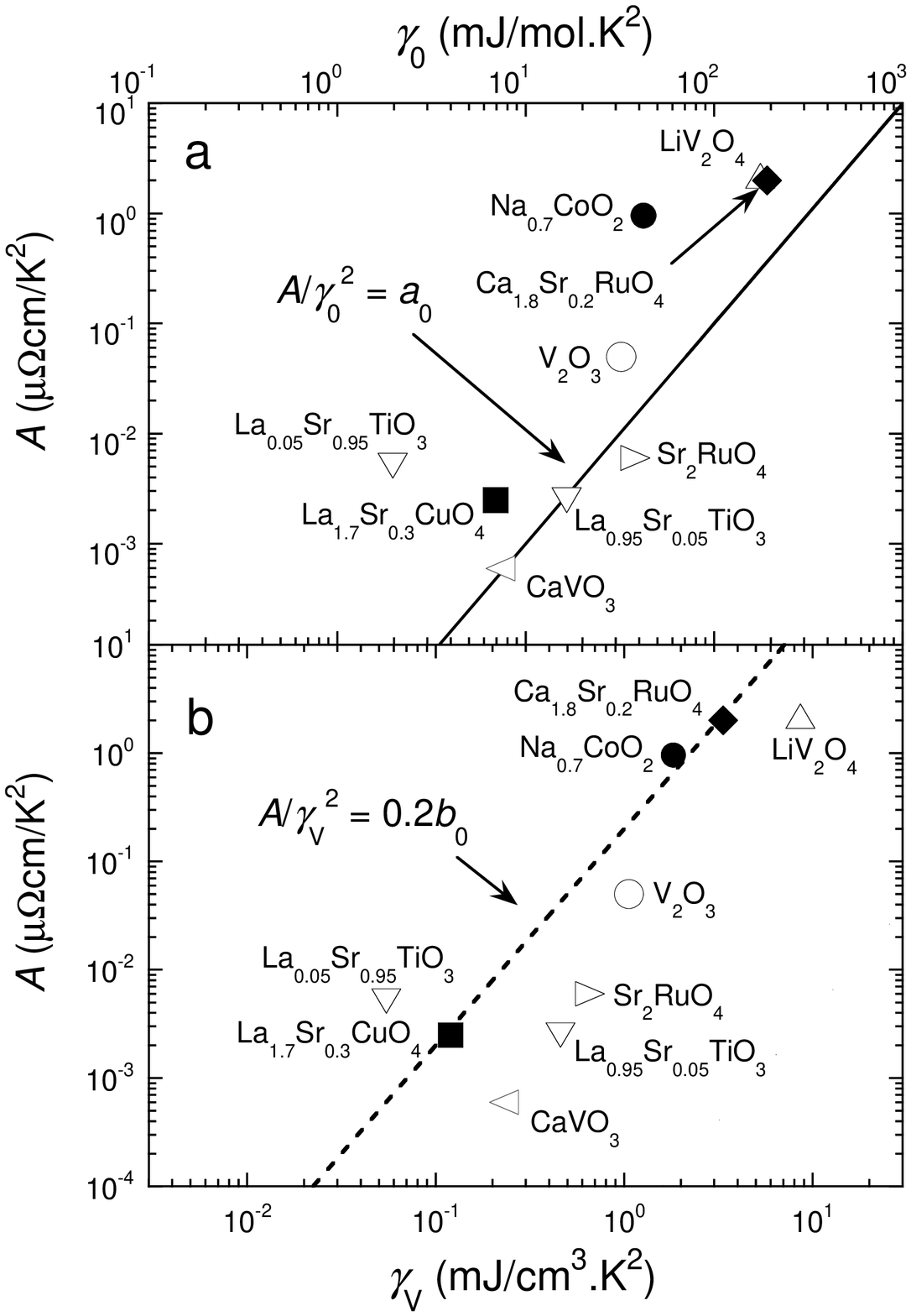}
\caption{a. Original Kadowaki-Woods plot ($A$ versus $\gamma_0^2$) for selected correlated oxides ($a_0$  = 10$^{-5}$ $\mu\Omega$cmK$^2$.mol$^2$/J$^2$).
b: $A$ versus $\gamma_V^2$ for the same compounds ($b_0$  = 1 $\mu\Omega$cmK$^2$.cm$^6$/J$^2$).}\label{fig1}
\end{figure}

\newpage
\noindent
\begin{figure}
\includegraphics[width=8.0cm,keepaspectratio=true]{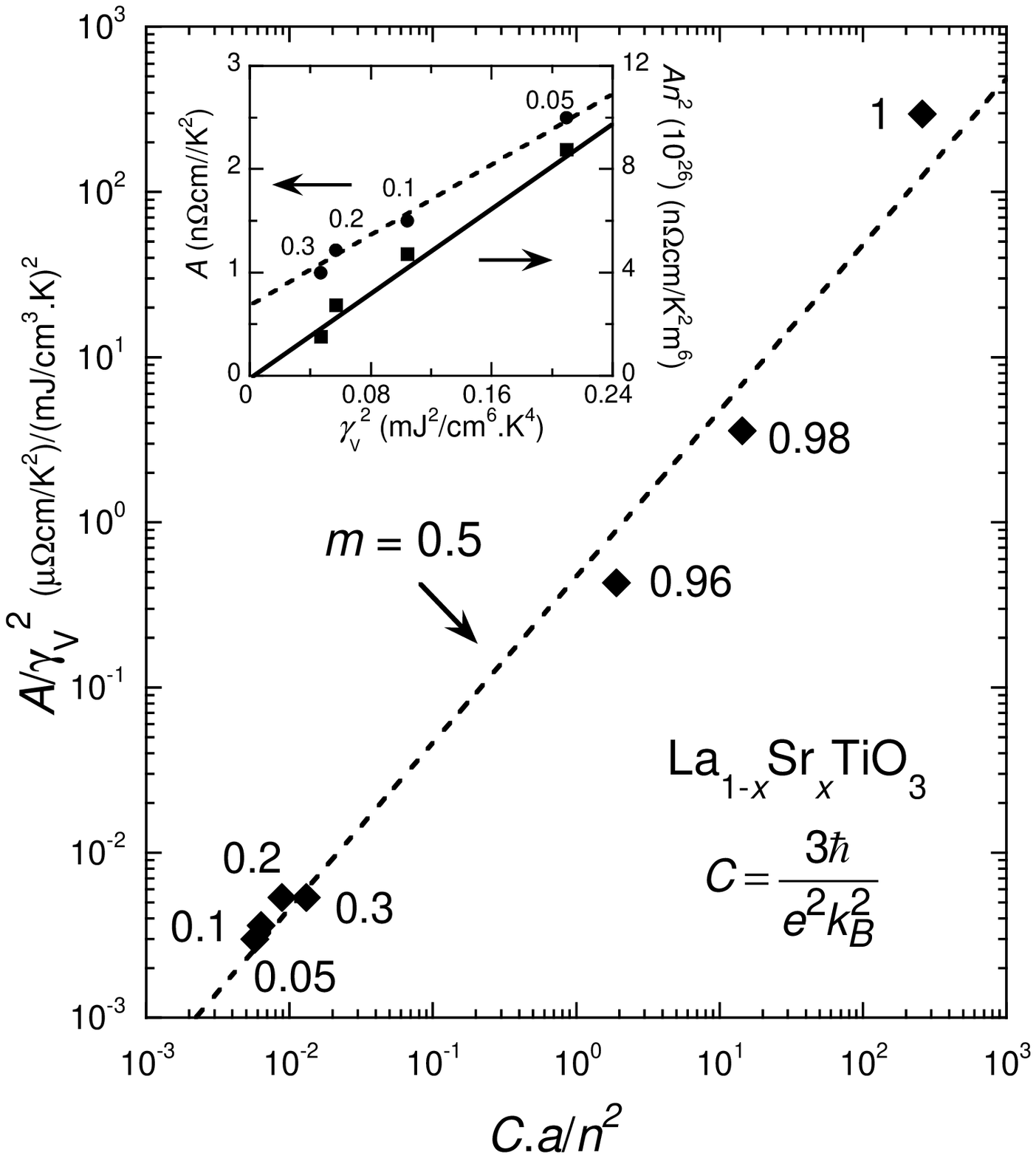}
\caption{$A$/$\gamma_V^2$ versus $C.a$/$n^2$ for La$_{1-x}$Sr$_x$TiO$_3$. The parameter $C$ is given in the figure. Each data point is labelled by
the appropriate $x$ value. Inset: Closed circles: $A$ versus $\gamma_V^2$ for La$_{1-x}$Sr$_x$TiO$_3$ near $x$ = 1. Black squares: $An^2$ versus
$\gamma_V^2$. The lines are guides to the eye.}\label{fig2}
\end{figure}

\end{document}